\documentclass[showpacs,preprintnumbers]{revtex4}
\usepackage{amsmath}
\usepackage{graphicx}
\usepackage{amsfonts}
\usepackage{amssymb}

\begin{document}
\textbf{The 4D geometric quantities versus the usual 3D quantities.}

\textbf{The resolution of Jackson's paradox}\medskip\bigskip

\qquad Tomislav Ivezi\'{c}

\qquad\textit{Ru\mbox
{\it{d}\hspace{-.15em}\rule[1.25ex]{.2em}{.04ex}\hspace{-.05em}}er Bo\v
{s}kovi\'{c} Institute, P.O.B. 180, 10002 Zagreb, Croatia}

\textit{\qquad ivezic@irb.hr\bigskip\bigskip}

In this paper we present definitions of different four-dimensional (4D)
geometric quantities (Clifford multivectors). New decompositions of the torque
$N$ and the angular momentum $M$ (bivectors) into 1-vectors $N_{s}$, $N_{t}$
and $M_{s}$, $M_{t}$ respectively are given. The torques $N_{s}$, $N_{t}$ (the
angular momentums $M_{s}$, $M_{t}$), taken \emph{together, }contain the same
physical information as the bivector $N$ (the bivector $M$). The usual
approaches that deal with the 3D quantities $\mathbf{E}$, $\mathbf{B}$,
$\mathbf{F}$, $\mathbf{L}$, $\mathbf{N}$, etc. and their transformations are
objected from the viewpoint of the invariant special relativity (ISR). In the
ISR it is considered that 4D geometric quantities are well-defined both
theoretically and \emph{experimentally} in the 4D spacetime. This is not the
case with the usual 3D quantities. It is shown that there is no apparent
electrodynamic paradox with the torque, and that the principle of relativity
is naturally satisfied, when the 4D geometric quantities are used instead of
the 3D quantities.\bigskip

\medskip\bigskip

\textbf{I. INTRODUCTION\bigskip}

It is almost generally accepted that the covariant quantities, e.g., the
covariant angular momentum four-tensor $M^{\mu\nu}$, the torque four-tensor
$N^{\mu\nu}$, the electromagnetic field strength tensor $F^{\alpha\beta}$,
etc. are only auxiliary mathematical quantities from which ``physical''
three-dimensional (3D) quantities, the angular momentum $\mathbf{L}$, the
torque $\mathbf{N}$, the electric and magnetic fields $\mathbf{E}$ and
$\mathbf{B}$, etc., are deduced. (The vectors in the 3D space will be
designated in bold-face.) The transformations of the 3D quantities are derived
from the Lorentz transformations (LT) of the corresponding covariant
quantities. (For such approaches see, e.g., [1-4].) However a geometric
approach to special relativity (SR) is recently developed, which exclusively
deals with 4D geometric quantities; it is called the invariant special
relativity (ISR). In the ISR one considers that the 4D geometric quantities
are well-defined both theoretically and \emph{experimentally} in the 4D
spacetime, and not, as usual, the 3D quantities. This geometric approach is
presented in [5-8] (tensor formalism, with tensors as geometric quantities)
and [9-13] (geometric algebra formalism). (See also [14] in which the
covariant 4-momentum of the electromagnetic field is expressed in terms of
4-vectors of the electric and magnetic fields.) It is shown in the mentioned
references that such geometric approach is in a complete agreement with the
principle of relativity and, what is the most important, with experiments, see
[7] (tensor formalism) and [9-13] (geometric algebra formalism).

In this paper the investigation with 4D geometric quantities will be done in
the geometric algebra formalism, see, e.g., [15, 16]. Physical quantities will
be represented by 4D geometric quantities, multivectors, that are defined
without reference frames, i.e., as absolute quantities (AQs) or, when some
basis has been introduced, they are represented as 4D coordinate-based
geometric quantities (CBGQs) comprising both components and a \emph{basis}.

In Sec. II we present the expressions for different 4D AQs and CBGQs; the
Lorentz force $K_{L}$ (1-vector) expressed in terms of the electromagnetic
field $F(x)$ (bivector), Eq. (\ref{lfk}), or in terms of the electric and
magnetic fields, $E$ and $B$ (1-vectors), Eq. (\ref{KEB}), the angular
momentum $M$ (bivector) and the torque $N$ (bivector), Eq. (\ref{MKN}). The
decomposition of $F(x)$ into 1-vectors $E$ and $B$ is given in (\ref{itf}).
The new decomposition of the torque $N$ into two 1-vectors, the
``space-space'' torque $N_{s}$ and the ``time-space'' torque $N_{t}$ is given
in (\ref{nls}); they \emph{together }contain the same physical information as
the bivector $N$. The similar decomposition of the angular momentum $M$ into
two 1-vectors $M_{s}$ and $M_{t}$ is presented in (\ref{lt}). $F(x)$ for a
charge $Q$ moving with constant velocity $u_{Q}$ (1-vector) is given in
(\ref{cvf}). The new expressions for the 1-vectors $E$ and $B$ for the same
case are given in (\ref{ec}).

In Sec. III we first discuss Jackson's [2] paradox with the 3D torque
$\mathbf{N}$. The paradox consists in the fact that there is a 3D torque
$\mathbf{N}$ and so a time rate of change of 3D angular momentum
($\mathbf{N}=d\mathbf{L}/dt$) in one inertial frame, but no 3D angular
momentum $\mathbf{L}^{\prime}$ and no 3D torque $\mathbf{N}^{\prime}$ in
another relatively moving inertial frame. Then it is shown that, contrary to
the general opinion, the transformations of the components of the 3D
quantities (e.g., Eq. (\ref{lk}) for the components of the 3D angular momentum
$\mathbf{L}$) drastically differ from the LT of the corresponding 4D
quantities (e.g., Eq. (\ref{ans}) for the components of the 4D angular
momentum $M_{s}$). Furthermore, a 4D geometric quantity, for example, 1-vector
$M_{s}$, is an invariant quantity under the LT, as can be seen from Eq.
(\ref{ms}); it is the same quantity for relatively moving inertial observers,
which can use different systems of coordinates. On the other hand the
corresponding 3D vector $\mathbf{L}$ in the inertial frame $S$ is completely
different than $\mathbf{L}^{\prime}$ in the relatively moving $S^{\prime}$
frame, as seen from Eq. (\ref{lc}). The same fundamental difference between
the 3D quantities and their transformations and the corresponding 4D geometric
quantities and their LT is discussed for some other quantities and equations
with them.

In Secs. IV - IV C the 4D geometric quantities from Sec. II are used to
resolve Jacson's paradox. First, in Sec. IV A, we considered the whole problem
using the bivector $N$ as an AQ and a CBGQ. It is shown that the paradox with
the 3D torque arises since all space-space components of $N$ as a CBGQ in the
$S^{\prime}$ frame, $N^{\prime ij}$, are zero but, as shown in (\ref{e}),
$N^{12}$ is different from zero in the $S$ frame. Since the components of the
3D torque are associated with the space-space components of $N$ this means
that $\mathbf{N}^{\prime}=\mathbf{0}$ but $\mathbf{N\neq0}$. From the point of
view of the ISR the fact that $\mathbf{N}^{\prime}\mathbf{\neq N}$ means that
$\mathbf{N}$ is not obtained by the LT from $\mathbf{N}^{\prime}$ and thus it
is not the same 4D quantity for observers in the $S^{\prime}$ and $S$ frames.
On the other hand when the 4D torque $N$ is used then it is shown that $N$ as
a CBGQ in $S$ (see (\ref{n})) is obtained by the LT from $N$ as a CBGQ in
$S^{\prime}$ (see (\ref{nc})); they represent the same 4D quantity in two
relatively moving inertial frames, $N$ ((\ref{nc})) $=$ $N$ ((\ref{n})). Hence
in the approach with the 4D torque $N$ the principle of relativity is
naturally satisfied and there is no paradox. In Secs. IV B and IV C we have
considered the same problem using the decomposition of $N$ into the
``space-space'' torque $N_{s}$ and the ``time-space'' torque $N_{t}$,
(\ref{nls}). Again the same result that there is no paradox is achieved. These
solutions with 4D geometric quantities can be simply applied to the
explanation of the Trouton-Noble experiment as shown in [12] and [13].

In Sec. V the conclusions are presented. \bigskip\medskip

\textbf{II. DEFINITIONS OF DIFFERENT 4D ABSOLUTE QUANTITIES\bigskip}

In this section, as already mentioned in the Introduction, we shall examine
different AQs and CBGQs. For simplicity and for easier understanding only the
standard basis \{$\gamma_{\mu};0,1,2,3$\} of orthonormal 1-vectors, with
timelike vector $\gamma_{0}$ in the forward light cone, will be used for
CBGQs. It is worth noting that the standard basis corresponds, in fact, to the
specific system of coordinates that we call Einstein's system of coordinates.
In Einstein's system of coordinates the standard, i.e., Einstein's
synchronization [17] of distant clocks and Cartesian space coordinates $x^{i}$
are used in the chosen inertial frame. However different systems of
coordinates are allowed in an inertial frame and they are all equivalent in
the description of physical phenomena. For example, in [5, 6] and the second
and third paper in [7], two very different, but physically completely
equivalent, systems of coordinates, Einstein's system of coordinates and the
system of coordinates with a nonstandard synchronization, the everyday (radio)
(``r'') synchronization, are exposed and exploited throughout the paper. In
order to treat different systems of coordinates on an equal footing we have
developed such form of the LT which is independent of the chosen system of
coordinates, including different synchronizations, [5, 6] (tensor formalism)
and [9] (Clifford algebra formalism). Furthermore in [6] we have presented the
transformation matrix that connects Einstein's system of coordinates with
another system of coordinates in the same reference frame. For the sake of
brevity and of clearness of the whole exposition, we shall only work with the
standard basis $\left\{  \gamma_{\mu}\right\}  $, but remembering that the
approach with 4D geometric quantities holds for any choice of basis.

Now let us write different physical quantities as 4D AQs and CBGQs. The
equations with them will be manifestly \emph{Lorentz} \emph{invariant}
\emph{equations}. The position 1-vector in the 4D spacetime is $x$. Then
$x=x(\tau)$ determines the history of a particle with proper time $\tau$ and
proper velocity $u=dx/d\tau$. The Lorentz force as a 4D AQ (1-vector) is
\begin{equation}
K_{L}=(q/c)F\cdot u, \label{lfk}%
\end{equation}
where $u$ is the velocity 1-vector of a charge $q$ (it is defined to be the
tangent to its world line). The bivector field $F(x)$ (i.e., the
electromagnetic field $F(x)$) for a charge $Q$ moving with constant velocity
$u_{Q}$ (1-vector) is
\begin{equation}
F(x)=kQ(x\wedge(u_{Q}/c))/\left|  x\wedge(u_{Q}/c)\right|  ^{3}, \label{cvf}%
\end{equation}
where $k=1/4\pi\varepsilon_{0}$, see [13] and references therein. (For the
charge $Q$ at rest, $u_{Q}/c=\gamma_{0}$.)

All AQs in Eq. (\ref{cvf}) can be written as CBGQs in some basis. We shall
write them in the standard basis $\{\gamma_{\mu}\}$. In the $\{\gamma_{\mu}\}$
basis $x=x^{\mu}\gamma_{\mu}$, $u_{Q}=u_{Q}^{\mu}\gamma_{\mu}$,
$F=(1/2)F^{\alpha\beta}\gamma_{\alpha}\wedge\gamma_{\beta}$; the basis
components $F^{\alpha\beta}$ are determined as $F^{\alpha\beta}=\gamma^{\beta
}\cdot(\gamma^{\alpha}\cdot F)=(\gamma^{\beta}\wedge\gamma^{\alpha})\cdot F$.
Every 4D CBGQ is invariant under the passive Lorentz transformations (LT); the
components transform by the LT and the basis by the inverse LT leaving the
whole CBGQ unchanged. (This is the reason for the name ISR.) The invariance of
some 4D CBGQ under the passive LT reflects the fact that such mathematical,
invariant, 4D geometric quantity represents \emph{the same physical quantity}
for relatively moving inertial observers. Due to the invariance of any 4D CBGQ
under the passive LT it will hold that, e.g., $F=(1/2)F^{\mu\nu}\gamma_{\mu
}\wedge\gamma_{\nu}=(1/2)F^{\prime\mu\nu}\gamma_{\mu}^{\prime}\wedge
\gamma_{\nu}^{\prime}$, where $F^{\mu\nu}$ and $F^{\prime\mu\nu}$ are
components and $\gamma_{\mu}$ and $\gamma_{\mu}^{\prime}$ are the basis
1-vectors in two relatively moving inertial frames $S$ and $S^{\prime}$
respectively. (Of course one could also use another basis, e.g., the basis
$\{r_{\mu}\}$ with ``r'' synchronization, in which $F$ will be represented as
$F=(1/2)F_{r}^{\mu\nu}r_{\mu}\wedge r_{\nu}=(1/2)F_{r}^{\prime\mu\nu}r_{\mu
}^{\prime}\wedge r_{\nu}^{\prime}$, where the primed quantities are the
Lorentz transforms of the unprimed ones.) The use of CBGQs enables us to have
clearly and correctly defined \emph{concept of sameness} of a physical system
for different observers. In the ISR only quantities that do not change upon
the passive LT have an independent physical reality, both theoretically and
\emph{experimentally}. When the physical laws are written with such Lorentz
invariant quantities as in ISR then the principle of relativity is
automatically satisfied and there is no need to postulate it as in Einstein's
SR. It is worth noting that Einstein's [17] formulation of SR deals with
Lorentz contraction, dilatation of time and the usual transformations of the
3D vectors $\mathbf{E}$ and $\mathbf{B}$ (see, e.g., Eqs. (11.148) and
(11.149) in [1]). However, e.g., the rest length and the Lorentz contracted
length are not the same 4D quantity for relatively moving observers, since the
transformed length $L_{0}(1-\beta^{2})^{1/2}$ is different than the rest
length $L_{0}$. Rohrlich [18] named the Lorentz contraction and other
transformations which do not refer to the same 4D quantity as the ``apparent''
transformations (AT). Similar ideas are expressed by Gamba [19]. Both Rohrlich
[18] and Gamba [19] considered that the same quantity for relatively moving
frames is a covariantly defined quantity (components of tensors) that retain
the same form under the LT. But any covariant quantity, e.g., the
electromagnetic field strength tensor $F^{\alpha\beta}$, consists of
components (numbers) that are taken (implicitly) in some basis. It is true
that these components refer to the same tensor quantity, but they cannot be
equal since the bases are not included, e.g., $F^{\mu\nu}\neq F^{\prime\mu\nu
}$, where $F^{\prime\mu\nu}$ are Lorentz transformed components. When the 4D
geometric quantities are used, as in the ISR, the concept of sameness becomes
very clear since, as mentioned above, every 4D CBGQ is \emph{invariant} under
the passive LT. Using 4D geometric quantities and the concept of sameness we
have shown in [6] and [7] that not only the Lorentz contraction but the
dilatation of time as well are the AT. Recently a fundamental result is
achieved in [8, 10] and [11] (both in the tensor formalism and in the Clifford
algebra formalism). There it is proved that the usual transformations of the
3D $\mathbf{E}$ and $\mathbf{B}$ are also the AT; they markedly differ from
the LT of the 4D geometric quantities that represent the electric and magnetic
fields. This result indicates that, contrary to the general belief, which
prevails from Einstein's fundamental paper [17], the usual transformations of
$\mathbf{E}$ and $\mathbf{B}$ are not relativistically correct. Comparison
with experiments in [7] and [10-12] clearly showed that the approach to SR
with 4D geometric quantities, i.e., the ISR, is in a true agreement with all
considered experiments. That agreement is independent of the chosen frame and
of the chosen system of coordinates in it.

Now let us see how the bivector field $F$ can be decomposed. Usually the
spacetime split is used for the decomposition of $F$ into the electric and
magnetic fields that are represented by bivectors, see Eqs. (58)-(60) in
Hestenes' paper [15]. This means that Hestenes' decomposition is an observer
dependent decomposition; an observer independent quantity $F$ is decomposed
into observer dependent bivectors of the electric and magnetic fields.

Instead of using the observer dependent decomposition from [15,16] we shall
make an analogy with the tensor formalism [20] (see also [5, 6] and [8]) and
represent the electric and magnetic fields by 1-vectors $E$ and $B$ [9] that
are defined without reference frames, i.e., as AQs
\begin{align}
F  &  =(1/c)E\wedge v+(IB)\cdot v,\nonumber\\
E  &  =(1/c)F\cdot v,\quad B=-(1/c^{2})I(F\wedge v), \label{itf}%
\end{align}
where $I$ is the unit pseudoscalar. ($I$ is defined algebraically without
introducing any reference frame, as in [21], Sec. 1.2.) The velocity $v$ and
all other quantities entering into the relations (\ref{itf}) are AQs. That
velocity $v$ characterizes some general observer. We can say, as in tensor
formalism [20,6,8] that $v$ is the velocity (1-vector) of a family of
observers who measures $E$ and $B$ fields. Of course the relations for $E$ and
$B$, Eq. (\ref{itf}), hold for any observer; they are manifestly Lorentz
invariant equations. Note that
\begin{equation}
E\cdot v=B\cdot v=0, \label{ie}%
\end{equation}
which yields that only three components of $E$ and three components of $B$ are
independent quantities. The relations (\ref{itf}) (and (\ref{ie})) that
connect $F$ and 1-vectors $E$ and $B$ are explained in more detail in [9-11].
We also remark that a complete and consistent formulation of classical
electromagnetism with the bivector field $F$ as the primary quantity is
presented in [12].

(It is shown in [10,11] that one can use another equivalent decomposition of
$F$ into bivectors $E_{Hv}$ and $B_{Hv}$
\begin{align}
F  &  =E_{Hv}+cIB_{Hv}\mathbf{,\quad} E_{Hv}=(1/c^{2})(F\cdot v)\wedge
v,\nonumber\\
B_{Hv}  &  =-(1/c^{3})I[(F\wedge v)\cdot v],\quad IB_{Hv}=(1/c^{3})(F\wedge
v)\cdot v. \label{he}%
\end{align}
However we shall use the decomposition (\ref{itf}) into 1-vectors $E$ and $B$
since it is much simpler and closer to the usual formulation with the 3D
$\mathbf{E}$ and $\mathbf{B}$.)

The 1-vectors $E$ and $B$ for a charge $Q$ moving with constant velocity
$u_{Q}$ can be determined from (\ref{itf}) and the expression for the bivector
field $F$ (\ref{cvf}). They are
\begin{align}
E  &  =(D/c^{2})[(x\wedge u_{Q})\cdot v]\nonumber\\
B  &  =(-D/c^{3})I(x\wedge u_{Q}\wedge v), \label{ec}%
\end{align}
where $D=kQ/\left|  x\wedge(u_{Q}/c)\right|  ^{3}$. Note that $B$ in
(\ref{ec}) can be expressed in terms of $E$ as
\begin{equation}
B=(1/c^{3})I(u_{Q}\wedge E\wedge v). \label{bi}%
\end{equation}
When the world lines of the observer and the charge $Q$ coincide, $u_{Q}=v$,
then (\ref{ec}) yields that $B=0$ and only an electric field (Coulomb field) remains.

The Lorentz force can be written in terms of 4D AQs, 1-vectors $E$ and $B$, as
[9,10]
\begin{equation}
K_{L}=(q/c)\left[  (1/c)E\wedge v+(IB)\cdot v\right]  \cdot u. \label{KEB}%
\end{equation}
Particularly from the definition of the Lorentz force $K_{L}$ and the relation
$E=(1/c)F\cdot v$ (from (\ref{itf})) it follows that the Lorentz force
ascribed by an observer comoving with a charge, $u=v$, is purely electric
$K_{L}=qE$. When $K_{L}$ is written as a CBGQ in $S$ and in the $\{\gamma
_{\mu}\}$ basis it is given as
\begin{equation}
K_{L}=(q/c^{2})[(v^{\nu}u_{\nu})E^{\mu}+\widetilde{\varepsilon}_{\ \nu\rho
}^{\mu}u^{\nu}cB^{\rho}-(E^{\nu}u_{\nu})v^{\mu}]\gamma_{\mu}, \label{lo}%
\end{equation}
where $\widetilde{\varepsilon}_{\mu\nu\rho}\equiv\varepsilon_{\lambda\mu
\nu\rho}v^{\lambda}$ is the totally skew-symmetric Levi-Civita pseudotensor
induced on the hypersurface orthogonal to $v$.

When the force, e.g., the Lorentz force $K_{L}$ (\ref{KEB}), is known we can
solve the equation of motion, Newton's second law, written as
\begin{equation}
K=dp/d\tau,\ p=mu, \label{eqmot}%
\end{equation}
where $p$ is the proper momentum (1-vector); $p$ as a CBGQ in $S$ and in the
$\{\gamma_{\mu}\}$ basis is $p=p^{\nu}\gamma_{\nu}$, $p^{\nu}=(\gamma
_{u}mc,\gamma_{u}p_{x},\gamma_{u}p_{y},\gamma_{u}p_{z})$, where $p_{x,y,z}$
are the components of the 3D momentum $\mathbf{p=}m\mathbf{u}$, $\gamma
_{u}=(1-\beta_{u}^{2})^{-1/2}$, $\beta_{u}=\left|  \mathbf{u}\right|  /c.$

Furthermore the angular momentum $M$ (bivector), the torque $N$ (bivector) for
the force $K$ and manifestly Lorentz invariant equation connecting $M$ and $N$
are defined as
\begin{align}
M &  =x\wedge p,\quad N=x\wedge K;\nonumber\\
N &  =dM/d\tau.\label{MKN}%
\end{align}
When $M$ and $N$ are written as CBGQs in the $\{\gamma_{\mu}\}$ basis they
become
\begin{align}
M &  =(1/2)M^{\mu\nu}\gamma_{\mu}\wedge\gamma_{\nu},\ M^{\mu\nu}=m(x^{\mu
}u^{\nu}-x^{\nu}u^{\mu}),\nonumber\\
N &  =(1/2)N^{\mu\nu}\gamma_{\mu}\wedge\gamma_{\nu},\ N^{\mu\nu}=x^{\mu}%
K^{\nu}-x^{\nu}K^{\mu}.\label{mn}%
\end{align}
We see that the components $M^{\mu\nu}$ from (\ref{mn}) are identical to the
covariant angular momentum four-tensor given by Eq. (A3) in Jackson's paper
[2]. However $M$ and $N$ from (\ref{MKN}) are 4D geometric quantities, the 4D
AQs, which are independent of the chosen reference frame and of the chosen
system of coordinates in it, whereas the components $M^{\mu\nu}$ and
$N^{\mu\nu}$ that are used in the usual covariant approach, e.g., Eq. (A3) in
[2], are coordinate quantities, the numbers obtained in the specific system of
coordinates, i.e., in the $\{\gamma_{\mu}\}$ basis. Notice that, in contrast
to the usual covariant approach, $M$ and $N$ from (\ref{mn}) are also 4D
geometric quantities, the 4D CBGQs, which contain both components and a
\emph{basis}, here bivector basis $\gamma_{\mu}\wedge\gamma_{\nu}$.

In the same way as $F$ is decomposed in (\ref{itf}) into 1-vectors $E$ and $B
$ and the unit time-like 1-vector $v/c$ we can decompose the bivector $N$
defined either as 4D AQs (by the relation (\ref{MKN})) or as 4D CBGQ (by
equation (\ref{mn}) into two 1-vectors $N_{s}$ and $N_{t}$
\begin{align}
N  &  =(v/c)\wedge N_{t}+(v/c)\cdot(N_{s}I),\nonumber\\
N_{t}  &  =(v/c)\cdot N,\quad N_{s}=I(N\wedge v/c), \label{nls}%
\end{align}
with the condition
\begin{equation}
N_{s}\cdot v=N_{t}\cdot v=0; \label{cs}%
\end{equation}
only three components of $N_{s}$ and three components of $N_{t}$ are
independent since $N$ is antisymmetric. Here, as in (\ref{itf}), $v$ is the
velocity (1-vector) of a family of observers who measures $N$. All quantities
in (\ref{nls}) are 4D AQs. It is worth noting that the introduction of $N_{s}$
and $N_{t}$ and the decomposition of the bivector $N$ into 1-vectors $N_{s}$
and $N_{t}$, equations (\ref{nls}) and (\ref{cs}), are not earlier mentioned
in the literature, as I am aware. When $N_{s}$ and $N_{t}$ are written as
CBGQs in the $\{\gamma_{\mu}\}$ basis they become
\begin{equation}
N_{s}=(1/2c)\varepsilon^{\alpha\beta\mu\nu}N_{\alpha\beta}v_{\mu}\gamma_{\nu
},\ N_{t}=(1/c)N^{\mu\nu}v_{\mu}\gamma_{\nu}. \label{ngs}%
\end{equation}

It is seen from (\ref{ngs}) that in the frame of ``fiducial'' observers, in
which the observers who measure $N_{s}$ and $N_{t}$ are at rest, and in the
$\{\gamma_{\mu}\}$ basis, $v^{\mu}=(c,0,0,0)$, $N_{s}^{0}=N_{t}^{0}=0$ and
\emph{only the spatial components} $N_{s}^{i}$ and $N_{t}^{i}$ remain
\begin{align}
N_{s}^{0}  &  =0,N_{s}^{i}=(1/2)\varepsilon^{0jki}N_{jk},\ N_{t}^{0}%
=0,\ N_{t}^{i}=N^{0i},\nonumber\\
N_{s}^{1}  &  =N^{23}=x^{2}K_{L}^{3}-x^{3}K_{L}^{2},\ N_{s}^{2}=N^{31}%
,\ N_{s}^{3}=N^{12}. \label{l1}%
\end{align}

Thus in our approach the torque in the 4D spacetime is the bivector $N$
defined either as 4D AQs (by the relation (\ref{MKN})) or as 4D CBGQ (by
equation (\ref{mn}). From the bivector $N$ we have constructed two 1-vectors,
the ``space-space'' torque $N_{s}$ and the ``time-space'' torque $N_{t}$ (the
relation (\ref{nls}) with the condition (\ref{cs})), which \emph{together
}contain the same physical information as the bivector $N$. Hence both $N_{s}$
and $N_{t}$ are 4D torques which \emph{taken together} are equivalent to the
4D torque, the bivector $N$.

The whole discussion with the torque can be completely repeated for the
angular momentum replacing $N$, $N_{s}$ and $N_{t}$ by $M$, $M_{s}$ and
$M_{t}$. Thus we have%

\begin{align}
M &  =(v/c)\wedge M_{t}+(v/c)\cdot(M_{s}I),\nonumber\\
M_{t} &  =(v/c)\cdot M,\quad M_{s}=I(M\wedge v/c),\label{lt}%
\end{align}
whith the condition
\begin{equation}
M_{s}\cdot v=M_{t}\cdot v=0.\label{cn}%
\end{equation}
1-vectors $M_{s}$ and $M_{t}$ correspond to $\mathbf{L}$ and $\mathbf{K}$ from
[2] respectively in the usual 3D picture. It has to be remarked that,
according to my knowledge, the relations (\ref{lt}) and (\ref{cn}) were not
earlier mentioned in the literature. It is usually considered that only
$M_{s}$ is angular momentum, see, e.g., Ludvigsen's book, [20] Sec. 8.3 (with
tensors as geometric quantities). In our approach both $M_{s}$ and $M_{t}$ are
angular momentums, which have to be treated on an equal footing. They contain
the same physical information as the bivector $M$ \emph{only when they are
taken together.}

When $M_{s}$ and $M_{t}$ are written as CBGQs in the $\{\gamma_{\mu}\}$
basis.they become
\begin{equation}
M_{s}=(1/2c)\varepsilon^{\alpha\beta\mu\nu}M_{\alpha\beta}v_{\mu}\gamma_{\nu
},\ M_{t}=(1/c)M^{\mu\nu}v_{\mu}\gamma_{\nu}. \label{lg}%
\end{equation}

The representation of the angular momentum with two 1-vectors $M_{s}$ and
$M_{t}$ will surely have important consequences in the quantum theory and in
the quantum field theory.

Instead of using decompositions of $N$ and $M$ into 1-vectors $N_{s}$, $N_{t}
$ and $M_{s}$, $M_{t}$ we could decompose them into bivectors in a complete
analogy with the decomposition of $F$ into the bivectors $E_{Hv}$ and $B_{Hv}
$ (\ref{he}), but it will not be done here.\medskip\bigskip

\textbf{III. JACKSON'S\ PARADOX. THE 3D QUANTITIES AND }

\textbf{THEIR APPARENT TRANSFORMATIONS\bigskip\ }

Having discussed different 4D geometric quantities we now examine the
corresponding 3D quantities and their transformations. A nice example from the
recent literature will help to better understand the essential difference
between the approach with 4D geometric quantities and the usual approach with
3D quantities.

In a recent paper Jackson [2] discussed the apparent paradox of different
mechanical equations for force and torque governing the motion of a charged
particle in different inertial frames. Two inertial frames $S$ (the laboratory
frame) and $S^{\prime}$ (the moving frame) are considered (they are $K$ and
$K^{\prime}$ respectively in Jackson's notation). In $S^{\prime}$ a particle
of charge $q$ and mass $m$ experiences only the radially directed electric
force caused by a point charge $Q$ fixed permanently at the origin.
Consequently both the angular momentum $\mathbf{L}^{\prime}$ and the torque
$\mathbf{N}^{\prime}$ are zero in $S^{\prime}$, see Fig. 1(a) in [2]. In $S$
the charge $Q$ is in uniform motion and it produces both an electric field
$\mathbf{E}$ and a magnetic field $\mathbf{B}$. The existence of $\mathbf{B}$
in $S$ is responsible for the existence of the 3D magnetic force
$\mathbf{F}=q\mathbf{u}\times\mathbf{B}$ and this force provides a 3D torque
$\mathbf{N}$ ($\mathbf{N}=\mathbf{x}\times\mathbf{F}$) on the charged
particle, see Fig. 1(b) in [2]. Consequently a nonvanishing 3D angular
momentum of the charged particle changes in time in $S$, $\mathbf{N}%
=d\mathbf{L}/dt$. Thus there is a 3D torque and so a time rate of change of 3D
angular momentum in one inertial frame, but no 3D angular momentum and no 3D
torque in another. Jackson [2] considers that there is no paradox and that
such result is relativistically correct result. (It has to be mentioned that
exactly the same paradox appears in the Trouton-Noble experiment, see, e.g.,
[4] and references therein.)

In Sec. III in [2] Jackson discusses ``Lorentz transformations of the angular
momentum between frames.'' He starts with the usual covariant definition of
the angular momentum tensor $M^{\mu\nu}=x^{\mu}p^{\nu}-x^{\nu}p^{\mu}$, Eq.
(8) in [2]. Notice that the standard basis $\left\{  \gamma_{\mu}\right\}  $,
i.e., Einstein's system of coordinates, is implicit in that definition,
actually, the implicit basis is bivector basis as in (\ref{mn}). Then the
\emph{components} $L_{i}$ of the \emph{3D vector }$\mathbf{L}$ (which is
called the angular momentum) are identified with the space-space
\emph{components} of $M^{\mu\nu}$ and the \emph{components} $L_{t,i}$ of the
\emph{3D vector }$\mathbf{L}_{t}$ (for which a physical interpretation is not
given) are identified with the three time-space \emph{components} of
$M^{\mu\nu}$ (we denote Jackson's $K_{i}$ with $L_{t,i}$, $\mathbf{K}$ with
$\mathbf{L}_{t}$). (Note that instead of the 3D $\mathbf{L}$ and
$\mathbf{L}_{t}$ we are dealing with 4D geometric quantities, 1-vectors
$M_{s}$ and $M_{t}$, defined by (\ref{lt}) and (\ref{cn}).)

This is in a complete analogy with the way in which (see [1] Sec. 11.9) the
components of 3D vectors $\mathbf{B}$ and $\mathbf{E}$ are identified with the
space-space and the time-space \emph{components} respectively of $F^{\mu\nu}$
\begin{equation}
B_{i}=(1/2c)\varepsilon_{ikl}F^{lk},\quad E_{i}=F^{i0}. \label{bfe}%
\end{equation}
(It is worth noting that Einstein's fundamental work [22] is the earliest
reference on covariant electrodynamics and on the identification of components
of $F^{\alpha\beta}$ with the components of the 3D $\mathbf{E}$ and
$\mathbf{B.}$)

The mentioned identification for $L_{i}$ and $L_{t,i}$ is
\begin{equation}
L_{i}=(1/2)\varepsilon_{ikl}M^{kl},\quad L_{t,i}=M^{0i}. \label{el}%
\end{equation}
The relations (\ref{bfe}) and (\ref{el}) show that the components $L_{i}$
correspond to $-B_{i}$ and $L_{t,i}$ to $-E_{i}$. In (\ref{bfe}) and
(\ref{el}) the components of the 3D vectors $\mathbf{B}$, $\mathbf{E}$ and
$\mathbf{L}$, $\mathbf{L}_{t}$ respectively are written with lowered (generic)
subscripts, since they are not the spatial components of the 4D quantities.
This refers to the third-rank antisymmetric $\varepsilon$ tensor too. The
super- and subscripts are used only on the components of the 4D quantities.
The 3D vectors $\mathbf{L}$ and $\mathbf{L}_{t}$, as \emph{geometric
quantities in the 3D space,} are constructed multiplying the components
$M^{\mu\nu}$ of \emph{a} \emph{4D geometric quantity} $M$, (\ref{el}), by the
unit 3D vectors $\mathbf{i}$, $\mathbf{j}$, $\mathbf{k}$, e.g., $\mathbf{L=}%
M^{23}\mathbf{i}+M^{31}\mathbf{j}+M^{12}\mathbf{k}$. Such procedure clearly
shows that in the approach from [2] the physical reality is attributed to the
3D vector $\mathbf{L}$ (but what is with a physical interpretation for
$\mathbf{L}_{t}$) and not to the whole set of components $M^{\mu\nu}$, i.e.,
the 4D geometric quantity $M$. Note that exactly the same procedure is applied
to construct \emph{geometric quantities in the 3D space }$\mathbf{B}$ and
$\mathbf{E}$ from the components $F^{\mu\nu}$ and the unit 3D vectors
$\mathbf{i}$, $\mathbf{j}$, $\mathbf{k}$. The objections to such procedure for
the construction of $\mathbf{B}$ and $\mathbf{E}$ are considered in detail in,
e.g., [8, 10] and [11], and they apply in the same measure to the construction
of $\mathbf{L}$ and $\mathbf{L}_{t}$. Some of them are the following.

(i) The whole procedure is made in an inertial frame of reference with the
Einstein system of coordinates, i.e., the standard basis $\left\{  \gamma
_{\mu}\right\}  $. In another system of coordinates that is different than the
Einstein system of coordinates, e.g., differing in the chosen synchronization
(as it is the 'r' synchronization considered in [5-7]), the identification of
$E_{i}$ with $F^{i0},$ as in (\ref{bfe}) (and also for $B_{i}$), or $L_{t,i}$
with $M^{0i}$ in (\ref{el}), is impossible and meaningless.

(ii) Furthermore the components $E_{i},$ $B_{i}$and $L_{t,i}$, $L_{i}$, of the
3D vectors $\mathbf{E}$, $\mathbf{B}$ and $\mathbf{L}_{t}$, $\mathbf{L}$
respectively are determined from 4D quantities written in the standard basis
$\left\{  \gamma_{\mu}\right\}  .$ Hence when forming the geometric quantities
the components would need to be multiplied with the unit 1-vectors $\gamma
_{i}$ and not with the unit 3D-vectors.

It is considered in [2] that the relations (\ref{el}) hold both in $S^{\prime
}$, the rest frame of the charges $q$ and $Q$, and in $S$, the laboratory
frame, which then leads to the usual transformations of the components of the
3D vector $\mathbf{L}$ that are given by Eq. 11 in [2]. We write them as
\begin{equation}
L_{1}=L_{1}^{\prime},\ L_{2}=\gamma(L_{2}^{\prime}-\beta L_{t,3}^{\prime
}),\ L_{3}=\gamma(L_{3}^{\prime}+\beta L_{t,2}^{\prime}). \label{lk}%
\end{equation}
Note that the components $L_{i}$ in $S$ are expressed by \emph{the mixture of
components} $L_{i}^{\prime}$ \emph{and} $L_{t,i}^{\prime}$ from $S^{\prime}$.
It is clear from the usual transformations (\ref{lk}) that the components of
\emph{the 3D angular momentum do not vanish in the laboratory frame }$S$,
\emph{even if they do in} $S^{\prime}$. In this case $L_{3}$ is different from
zero due to contribution from $L_{t,2}^{\prime}$. Then Jackson [2] calculates
$dL_{3}/dt$, where $L_{3}$ is obtained ``via a Lorentz transformation,'' i.e.,
via the transformations (\ref{lk}) (note that in [2] the derivative of $L_{3}$
is relative to the coordinate time $t$ and not, as in (\ref{MKN}), relative to
the proper time $\tau$). It is shown in [2] that $dL_{3}/dt$ is $=N_{3}$,
where the torque $N_{3}$ ($N_{z}$ in Eq. (7) in [2]) is ``directly obtained
from the force equation in the laboratory.'' $N_{z}$ in Eq. (7) in [2] is
obtained using the force equation (Eq. (4) in [2]) with the 3D vectors
$\mathbf{p}$, $\mathbf{F}$,\textbf{\ }$\mathbf{E}$ and $\mathbf{B}$, but again
the derivative is relative to the coordinate time $t$. Jackson [2] finds the
consistency in both calculations and states: ``The time rate of change of the
particle's angular momentum obtained via a Lorentz transformation is equal to
the torque directly obtained from the force equation in the laboratory, as it
must.'' In our opinion what is found in [2] is that when using the 3D vectors
and their transformations (like Eq. (\ref{lk})) the paradox is always obtained
and, actually, the principle of relativity is violated.

Let us examine in more detail the transformations of components of the 3D
quantities and also of the 3D vectors. As already said both $\mathbf{L}%
^{\prime}$ and $\mathbf{L}$, as \emph{geometric quantities in the 3D space,}
are constructed multiplying the components $L_{i}^{\prime}$ and $L_{i}$ (given
by (\ref{lk})) by the unit 3D vectors $\mathbf{i}^{\prime}$, $\mathbf{j}%
^{\prime}$, $\mathbf{k}^{\prime}$ and $\mathbf{i}$, $\mathbf{j}$, $\mathbf{k}$
respectively. This gives another objection to the usual construction of
$\mathbf{L}^{\prime}$ and $\mathbf{L.}$

(iii) The components $L_{i}$ are determined by the transformations (\ref{lk}),
but there is no transformation which transforms the unit 3D vectors
$\mathbf{i}^{\prime}$, $\mathbf{j}^{\prime}$, $\mathbf{k}^{\prime}$ into the
unit 3D vectors $\mathbf{i}$, $\mathbf{j}$, $\mathbf{k}$. Hence it is not true
that the 3D vector $\mathbf{L=}L_{1}\mathbf{i}+L_{2}\mathbf{j}+L_{3}%
\mathbf{k}$ is obtained by the LT from the 3D vector $\mathbf{L}^{\prime
}\mathbf{=}L_{1}^{\prime}\mathbf{i}^{\prime}+L_{2}^{\prime}\mathbf{j}^{\prime
}+L_{3}^{\prime}\mathbf{k}^{\prime}$. Cosequently $\mathbf{L}$ and
$\mathbf{L}^{\prime}$ are not the same quantity for relatively moving inertial
observers, $\mathbf{L\neq L}^{\prime}$,
\begin{equation}
L_{1}\mathbf{\mathbf{i}+}L_{2}\mathbf{\mathbf{j}+}L_{3}\mathbf{\mathbf{k}\neq}%
L_{1}^{\prime}\mathbf{i}^{\prime}+L_{2}^{\prime}\mathbf{j}^{\prime}%
+L_{3}^{\prime}\mathbf{k}^{\prime}. \label{lc}%
\end{equation}
Thus contrary to the general opinion, the transformations (\ref{lk}) are not
the LT but the AT of the 3D $\mathbf{L}$. The same situation happens with the
transformations of the 3D $\mathbf{B}$ and $\mathbf{E}$ as explained in detail
in [8, 10] and [11].

On the other hand, as already mentioned in Sec. II, every 4D CBGQ is invariant
under the passive LT, which means that such 4D geometric quantity represents
the same physical quantity for relatively moving inertial observers. Hence, it
holds, e.g., $M_{t}=(1/c)M^{\mu\nu}v_{\mu}\gamma_{\nu}=(1/c)M^{\prime\mu\nu
}v_{\mu}^{\prime}\gamma_{\nu}^{\prime}$, where all primed quantities are
obtained by the LT from the unprimed ones. $M_{s}$ and $M_{t}$, written as 4D
CBGQs, transform under the LT as every 1-vector (as 4D CBGQ) transforms, which
means that the components $M_{s}^{\mu}$ transform again to $M_{s}^{\prime\mu}$
and similarly $M_{t}^{\mu}$ transform to $M_{t}^{\prime\mu}$; \emph{there is
no mixing of components}. Thus the equation corresponding to Eq. (11) in [2],
i.e., to Eq. (\ref{lk}), will be
\begin{equation}
M_{s}^{0}=\gamma(M_{s}^{\prime0}+\beta M_{s}^{\prime1}),\ M_{s}^{1}%
=\gamma(M_{s}^{\prime1}+\beta M_{s}^{\prime0}),\ M_{s}^{2,3}=M_{s}^{\prime
2,3}, \label{ans}%
\end{equation}
and the same for $M_{t}^{\mu}$. This is in a sharp contrast to the AT
(\ref{lk}) in which the transformed components $L_{i}$ are expressed by the
mixture of components $L_{k}^{\prime}$ and $L_{t,k}^{\prime}$. Furthermore
$M_{s}$ and $M_{t}$ are geometric quantities in the 4D spacetime since the
components $M_{s}^{\mu}$ and $M_{t}^{\mu}$ are multiplied by the unit
1-vectors $\gamma_{\mu}$, while, as we mentioned, the 3D angular momentum
$\mathbf{L}$ is formed multiplying the components $M^{\mu\nu}$ (i.e., $L_{i}$
determined by (\ref{el})) of a 4D geometric quantity $M$, by the unit 3D
vectors $\mathbf{i}$, $\mathbf{j}$, $\mathbf{k}$. Of course, it holds that
$M_{s}$ is the same quantity for observers in $S$ and $S^{\prime}$, which can
use different basis, e.g., $\left\{  \gamma_{\mu}\right\}  $, $\{r_{\mu}\}$
and so on. Thus
\begin{equation}
M_{s}=M_{s}^{\mu}\gamma_{\mu}=M_{s}^{\prime\mu}\gamma_{\mu}^{\prime}%
=M_{s,r}^{\mu}r_{\mu}=M_{s,r}^{\prime\mu}r_{\mu}^{\prime}=.., \label{ms}%
\end{equation}
where the primed quantities are the Lorentz transforms of the unprimed ones;
see the discussion at the beginning of Sec. II.

This is in a complete analogy with the fundamental difference between the AT
of the 3D $\mathbf{E}$ and $\mathbf{B}$ and the LT of 1-vectors $E$ and $B$
that are defined by (\ref{itf}); see, e.g., [10, 11], in which this
fundamental difference is exactly proved. There, it is also shown that the LT
of 1-vectors $E$ and $B$ are in a complete agreement with experiments on
motional emf and Faraday disk, while it is not the case with the AT of the 3D
$\mathbf{E}$ and $\mathbf{B}$. (Regarding the mentioned analogy, e.g., the AT
for the components of the 3D $\mathbf{B}$ are the same as the AT for $L_{i}$
that are given by (\ref{lk}); $L_{i}$, $L_{i}^{\prime}$ have to be replaced by
$B_{i}$, $B_{i}^{\prime}$ (components of the 3D $\mathbf{B}$, $\mathbf{B}%
^{\prime}$) and $L_{t,i}$, $L_{t,i}^{\prime}$ by $E_{i}$, $E_{i}^{\prime}$
(components of the 3D $\mathbf{E}$, $\mathbf{E}^{\prime}$). On the other hand
the LT for the components $B^{\mu}$ of the 1-vector $B$ are the same as the LT
(\ref{ans}), but $M_{s}^{\mu}$, $M_{s}^{\prime\mu}$ have to be replaced by
$B^{\mu}$, $B^{\prime\mu}$.)

The above consideration suggests that the transformations of other 3D
quantities are the AT as well. For example, the AT of the 3D torque
$\mathbf{N,}$ which are the same as (\ref{lk}), are found, e.g., in
Jefimenko's book [3] and given in [4] Eqs. (1)-(3). In Sec. 8 of [3], under
the title: ``From relativistic electromagnetism to relativistic mechanics,''
the AT of different 3D quantities are presented. Notice that the AT of the 3D
$\mathbf{E}$, $\mathbf{B}$, $\mathbf{L}$ and $\mathbf{N}$ may all be obtained
in the same way by the identification of the components of the 3D vectors with
the components of second-rank 4D tensors, i.e., bivectors in the 4D spacetime.

Furthermore the transformations of the 3D force $\mathbf{F}$ are also the AT;
they are given, e.g., by Eqs. (8-5.4)-(8-5.6) in [3] (or by Eqs. (1.53)-(1.55)
in [23])
\begin{equation}
F_{x}^{\prime}=[F_{x}-(\beta_{u}/c)(\mathbf{Fu})]/(1-(\beta_{u}u_{x}%
/c)),\ F_{y,z}^{\prime}=F_{y,z}/\gamma(1-(\beta_{u}u_{x}/c)),\label{fo}%
\end{equation}
where $\mathbf{u}$ is the 3D velocity of a particle. All previously mentioned
objections, (i) - (iii), regarding the construction of the 3D vectors, are
also at place here. The 3D forces $\mathbf{F}$, $\mathbf{F}^{\prime}$ are
constructed from the components $F_{x,y,z}$, $F_{x,y,z}^{\prime}$ (determined
by (\ref{fo})) and the unit 3D vectors $\mathbf{i}$, $\mathbf{j}$,
$\mathbf{k}$, and $\mathbf{i}^{\prime}$, $\mathbf{j}^{\prime}$, $\mathbf{k}%
^{\prime}$ respectively and $\mathbf{i}^{\prime}$, $\mathbf{j}^{\prime}$,
$\mathbf{k}^{\prime}$ are not obtained by any transformations from
$\mathbf{i}$, $\mathbf{j}$, $\mathbf{k}$. Particularly it is visible from
(\ref{fo}) that $\mathbf{F}^{\prime}\neq\mathbf{F}$; they do not refer to the
same quantity in the 4D spacetime and the transformations (\ref{fo}) are not
the LT but the AT. The same holds for the well-known transformations of the 3D
velocity $\mathbf{u}$ that are given, e.g., by equations (11.31) in [1], or by
equations (7-2.5)-(7-2.7) in [3].

From the ISR viewpoint, the correctly defined quantities in the 4D spacetime,
both theoretically and \emph{experimentally}, are 1-vectors $K=K^{\nu}%
\gamma_{\nu}$ and $u=u^{\nu}\gamma_{\nu}$, or in another basis $\{r_{\nu}\}$
these 1-vectors are $K=K_{r}^{\nu}r_{\nu}$ and $u=u_{r}^{\nu}r_{\nu}$. In
contrast to awkward transformations of the components of the 3D force
$\mathbf{F}$ (\ref{fo}) the LT of the components $K^{\nu}$ of the 1-vector $K$
are very simple (the LT (\ref{ans}) but with $K^{\mu}$, $K^{\prime\mu}$
replacing $M_{s}^{\mu}$, $M_{s}^{\prime\mu}$). When the components of the
4-force $K$ and of the 4-velocity $u$ are determined in the standard basis
$\{\gamma_{\nu}\}$ then they can be expressed in terms of components of the 3D
force $\mathbf{F}$, $(F_{x},F_{y},F_{z})$, and of the 3D velocity $\mathbf{u}%
$, $(u_{x},u_{y},u_{z})$. They are $K^{\nu}=(\gamma_{u}\mathbf{Fu}%
/c,\gamma_{u}F_{x},\gamma_{u}F_{y},\gamma_{u}F_{z})$ and $u^{\nu}=(\gamma
_{u}c,\gamma_{u}u_{x},\gamma_{u}u_{y},\gamma_{u}u_{z})$, where $\gamma
_{u}=(1-\left|  \mathbf{u}\right|  ^{2}/c^{2})^{-1/2}$. We see that, in
general, the spatial components $K^{i}$, $u^{i}$ differ from the components of
the 3D quantities $\mathbf{F}$, $\mathbf{u}$. Only in the case when the
considered particle is at rest, i.e., $u_{x,y,z}=0$, $\gamma_{u}=1$ and
consequently $u^{\nu}=(c,0,0,0)$, then $K^{\nu}$ will be exclusively
determined with the components $F_{x,y,z}$, i.e., $K^{\nu}=(0,F_{x}%
,F_{y},F_{z})$. However even in that case $u^{\nu}$ and $K^{\nu}$ are the
components of the \emph{4D geometric quantities }$u=u^{\nu}\gamma_{\nu}$ and
$K=K^{\nu}\gamma_{\nu}$ in the $\{\gamma_{\nu}\}$ basis and not the components
of some \emph{3D geometric quantities} $\mathbf{u}$ and $\mathbf{F}$, see also
the discussion in [10] Sec. 3.2. This discussion additionaly shows that the
transformations (\ref{fo}) have nothing in common with the LT of the 1-vector
$K$. 

It is generally accepted, e.g., [1, 3, 23], that the ``relativistic''
equations of motion have the same form in two relatively moving inertial
frames $S$ and $S^{\prime}$
\begin{align}
\mathbf{F}  &  =d\mathbf{p}/dt,\ \mathbf{p=}m\gamma_{u}\mathbf{u}\nonumber\\
\mathbf{F}^{\prime}  &  =d\mathbf{p}^{\prime}/dt^{\prime}, \label{f2}%
\end{align}
see, for example, Eqs. (1.39) and (1.40) in [23], or Sec. 12.2 and 12.4 (with
the Lorentz force) in [1].

In Einstein's formulation [17] of SR the principle of relativity is a
fundamental postulate that is supposed to hold for all physical laws including
those expressed by 3D quantities, e.g., the Maxwell equations with the 3D
$\mathbf{E}$ and $\mathbf{B}$; Einstein [17] used that postulate to derive the
transformations of the 3D $\mathbf{E}$ and $\mathbf{B}$. It is proved in [11],
both in the geometric algebra and tensor formalisms, that the usual Maxwell
equations with the 3D $\mathbf{E}$ and $\mathbf{B}$ change their form under
the LT and thus that they are not covariant under the LT. This result
explicitly shows that the principle of relativity does not hold for physical
laws expressed by 3D quantities (a fundamental achievement). The results from
this section also reveal that a 3D quantity cannot correctly transform under
the LT, which means that it does not have an independent physical reality in
the 4D spacetime; it is not the same quantity for relatively moving observers
in the 4D spacetime. Hence it is not true that Eqs. (\ref{f2}) are the
relativistic equations of motion since the primed 3D quantities are not
obtained by the LT from the unprimed ones, but they are obtained in terms of
the AT for the 3D force $\mathbf{F}$ (\ref{fo}) and the 3D momentum
$\mathbf{p}$, i.e., the 3D velocity $\mathbf{u}$, Eq. (11.31) in [1]. Instead
of Eqs. (\ref{f2}) one has to use equation of motion with 4D geometric
quantities (\ref{eqmot}). 

Similarly it is generally accepted in usual approaches that the Lorentz force
law remains of the same form in relatively moving inertial frames $S$ and
$S^{\prime}$%
\begin{equation}
S;\ \mathbf{F=}q\mathbf{E}+q\mathbf{u}\times\mathbf{B,\quad} S^{\prime
};\ \mathbf{F}^{\prime}\mathbf{=}q\mathbf{E}^{\prime}+q\mathbf{u}^{\prime
}\times\mathbf{B}^{\prime}, \label{f5}%
\end{equation}
where all primed quantities are considered to be obtained by the LT from the
unprimed ones. For example, it is argued, e.g., in [3] Sec. 8: ``This law does
not depend on the inertial reference frame in which $q$, $\mathbf{u}$,
$\mathbf{E}$, and $\mathbf{B}$ are measured.'' The same assertions about the
form of the Lorentz force law in two inertial frames can be found in [23],
Eqs. (6.42) and (6.43). There, this form invariance of the Lorentz force,
together with the AT of the 3D force $\mathbf{F}$ (\ref{fo}) and the AT of the
3D velocity $\mathbf{u}$, Eqs. (1.26)-(1.28) in [23], are used to derive the
AT for the 3D $\mathbf{E}$ and $\mathbf{B}$. However the above discussion
clearly shows that the 3D quantities in $S^{\prime}$ are not obtained by the
LT from the corresponding 3D quantities in $S$ than by the use of the AT.
Therefore the Lorentz force law has to be written by means of 4D geometric
quantities, e.g., Eq. (\ref{lfk}) with bivector field $F$, or Eq. (\ref{KEB})
with 1-vectors $E$ and $B$ and, of course, so defined $K_{L}$ is an invariant
quantity under the LT. \medskip\bigskip

\textbf{IV. THE RESOLUTION\ OF\ JACKSON'S\ PARADOX USING 4D TORQUES\bigskip\ }

Instead of dealing with 3D quantities $\mathbf{E}$, $\mathbf{B}$, $\mathbf{L}$
and $\mathbf{N}$ and their AT as in [2] we shall examine Jackson's paradox
using the expressions for the 4D geometric quantities from Sec. II. First we
write $N$ from (\ref{MKN}) using the expression (\ref{lfk}) for $K_{L}$ and
(\ref{cvf}) for $F$. Then $N$ becomes
\begin{equation}
N=(Dq/c^{2})(u\cdot x)(u_{Q}\wedge x), \label{AN}%
\end{equation}
where $D$ is already defined in (\ref{ec}), $D=kQ/\left|  x\wedge
(u_{Q}/c)\right|  ^{3}$. This is the most general expression for the
considered torque $N$ written as an AQ. Then $N_{s}$ and $N_{t}$ are
determined from (\ref{nls}) and (\ref{AN}) as
\begin{align}
N_{s}  &  =(Dq/c^{3})(u\cdot x)I(x\wedge v\wedge u_{Q}),\nonumber\\
N_{t}  &  =(Dq/c^{3})(u\cdot x)[(x\wedge u_{Q})\cdot v]. \label{n1}%
\end{align}
Comparison with (\ref{ec}) shows that $N_{s}$ and $N_{t}$ can be expressed in
terms of $B$ and $E$ as
\begin{align}
N_{s}  &  =q(u\cdot x)B,\nonumber\\
N_{t}  &  =(q/c)(u\cdot x)E. \label{n2}%
\end{align}
As already said, in connection with (\ref{ec}), when $u_{Q}=v$ then $B=0$. The
relations (\ref{n1}) and (\ref{n2}) reveal that in that case $N_{s}=0$ as
well.\bigskip\medskip

\textbf{A. }$N$ \textbf{as a CBGQ in }$S^{\prime}$ \textbf{and }$S$
\textbf{frames}\bigskip

Let us now write all AQs from (\ref{AN}) as CBGQs in $S^{\prime}$, the rest
frame of the charge $Q$, in which $u_{Q}=c\gamma_{0}^{\prime}$. Then
$N=(Dq/c)(u\cdot x)(\gamma_{0}^{\prime}\wedge x)$, and in the $\{\gamma_{\mu
}^{\prime}\}$ basis it is explicitly given as
\begin{align}
N  &  =(1/2)N^{\prime\mu\nu}\gamma_{\mu}^{\prime}\wedge\gamma_{\nu}^{\prime
}=N^{\prime01}(\gamma_{0}^{\prime}\wedge\gamma_{1}^{\prime})+N^{\prime
02}(\gamma_{0}^{\prime}\wedge\gamma_{2}^{\prime}),\nonumber\\
N^{\prime01}  &  =(Dq/c)(u^{\prime\mu}x_{\mu}^{\prime})x^{\prime1}%
,\ N^{\prime02}=(Dq/c)(u^{\prime\mu}x_{\mu}^{\prime})x^{\prime2}. \label{nc}%
\end{align}
The components $x^{\prime\mu}$ are $x^{\prime\mu}=(x^{\prime0}=ct^{\prime
},x^{\prime1},x^{\prime2},0)$ where $x^{\prime1}=r^{\prime}\cos\theta^{\prime
}$, $x^{\prime2}=r^{\prime}\sin\theta^{\prime}$. In $S^{\prime}$ the velocity
1-vector of the charge $q$ ($u=dx/d\tau$) at any $t^{\prime}$ is
$u=u^{\prime\mu}\gamma_{\mu}^{\prime}$, where $u^{\prime\mu}=dx^{\prime\mu
}/d\tau=(u^{\prime0},u^{\prime1},u^{\prime2},0)$. The components $N^{^{\prime
}\mu\nu}$ ($N^{\prime\mu\nu}=x^{\prime\mu}K_{L}^{\prime\nu}-x^{\prime\nu}%
K_{L}^{\prime\mu}$) that are different from zero are only $N^{\prime01}$ and
$N^{\prime02}$.

It can be easily seen that \emph{all} $N^{\prime\alpha\beta}$ \emph{are zero
in the} $S^{\prime}$ \emph{frame} when it is supposed that at $t^{\prime}=0$
the charge $q$ is still at rest, i.e., $u^{\prime\mu}=(c,0,0,0)$. From the
invariance of any 4D CBGQ under the passive LT it follows that at $t^{\prime
}=0$ the whole $N$ is zero not only in $S^{\prime}$ but in the laboratory
frame $S$ as well. This case explicitly refers to the Trouton-Noble paradox as
discussed in [12, 13].

Similarly in order to find the torque $N$ in the $S$ frame we write all AQs
from (\ref{AN}) as CBGQs in $S$ and in the $\{\gamma_{\mu}\}$ basis. In $S$
the charge $Q$ is moving with velocity $u_{Q}=\gamma_{Q}c\gamma_{0}+\gamma
_{Q}\beta_{Q}c\gamma_{1}$, where $\beta_{Q}=\left|  \mathbf{u}_{Q}\right|  /c$
and $\gamma_{Q}=(1-\beta_{Q}^{2})^{-1/2}$. Another way to find $N$ in $S$ is
to make the LT of $N$ in $S^{\prime}$ ((\ref{nc})). The result is
\begin{equation}
N=(1/2)N^{\mu\nu}\gamma_{\mu}\wedge\gamma_{\nu}=N^{01}\gamma_{0}\wedge
\gamma_{1}+N^{02}\gamma_{0}\wedge\gamma_{2}+N^{12}\gamma_{1}\wedge\gamma
_{2},\label{g}%
\end{equation}
where $N^{01}=N^{\prime01}$, $N^{02}=\gamma_{Q}N^{\prime02}$ and
$N^{12}=\gamma_{Q}\beta_{Q}N^{\prime02}$. When $N$ is written explicitly in
terms of quantities in $S$ it becomes
\begin{align}
N &  =(Dq/c)(u^{\mu}x_{\mu})[\gamma_{Q}(x^{1}-\beta_{Q}x^{0})(\gamma_{0}%
\wedge\gamma_{1})\nonumber\\
&  +\gamma_{Q}x^{2}(\gamma_{0}\wedge\gamma_{2})+\beta_{Q}\gamma_{Q}%
x^{2}(\gamma_{1}\wedge\gamma_{2})].\label{n}%
\end{align}
We see that in the laboratory frame, where the charge $Q$ is moving, the
components $N^{\mu\nu}$ that are different from zero are not only $N^{01}$ and
$N^{02}$ but also $N^{12}=\beta_{Q}N^{02}$.

Of course, we could start with the 4D angular momentum $M=x\wedge p$ and
calculate it in both frames $S^{\prime}$ and $S$. Then using the relation
$N=dM/d\tau$ one can again find the same expressions (\ref{nc}) and (\ref{n})
for $N$.

The component $N^{12}$ can be written in the form similar to the expression
for $N_{z}$, Eq. (7) in [2]. Thus
\begin{equation}
N^{12}=\beta_{Q}ctK_{L}^{2}-\beta_{Q}yK_{L}^{0}=\beta_{Q}ctK_{L}%
^{2}-(q/c)\beta_{Q}y(F^{0i}u_{i}). \label{e}%
\end{equation}
$K_{L}^{2}$ can be determined from (\ref{lfk}) and it is $K_{L}^{2}%
=(q/c)(F^{20}u_{0}+F^{21}u_{1})$. Comparing $N_{z}$ from Eq. (7) in [2] and
our $N^{12}$ from (\ref{e}) we see that instead of the components of the 3D
Lorentz force $\mathbf{F}$, we have the components $K_{L}^{2}$ and $K_{L}^{0}$
of the 4D Lorentz force $K_{L}$, which is defined by (\ref{lfk}).

This form for $N^{12}$ clearly shows the essential difference between our
approach and the usual approach, e.g., [2]. The paradox with the 3D torque
arises since all $N^{\prime ij}$ are zero but, according to (\ref{e}),
$N^{12}$ is different from zero in the $S$ frame, which means that
$\mathbf{N}^{\prime}=\mathbf{0}$ but $\mathbf{N\neq0}$. From the point of view
of the ISR the fact that $\mathbf{N}^{\prime}\mathbf{\neq N}$ means that
$\mathbf{N}$ is not obtained by the LT from $\mathbf{N}^{\prime}$ and thus it
is not the same 4D quantity for observers in the $S^{\prime}$ and $S$ frames.
In contrast to the usual approach [2] with 3D quantities, $N$ determined by
(\ref{n}) is obtained by the LT from $N$ given by (\ref{nc}); they represent
the same 4D quantity in two relatively moving inertial frames, $N$
((\ref{nc})) $=$ $N$ ((\ref{n})). Hence the principle of relativity is
naturally satisfied in our geometric approach and there is not any paradox.

This consideration indicates that in order to check the validity of the above
relations with 4D quantities, and thus the validity of the principle of
relativity and, more generally, of the SR, the measurements must be changed
relative to the usual measurements of the 3D quantities. For the 4D quantities
the experimentalists have to measure \emph{all components} of $N$ and $M$ in
both frames $S^{\prime}$ and $S$. The observers in $S^{\prime}$ and $S$ are
able to compare only such complete set of data which corresponds to the
\emph{same} 4D geometric quantity. Such point of view is illustrated in much
more detail in [7].\bigskip\medskip

\textbf{B. }$N_{s}$ \textbf{and} $N_{t}$ \textbf{as CBGQs in }$S^{\prime}$
\textbf{and }$S$ \textbf{frames. }$S^{\prime}$ \textbf{is }

\textbf{the frame} \textbf{of ``fiducial'' observers}\bigskip

Let us now make the same consideration as in Sec. IV A, but with $N_{s}$ and
$N_{t}$ as CBGQs. From the relations (\ref{nls}) and (\ref{cs}) we see that
1-vectors $N_{s}$ and $N_{t}$ are not uniquely determined by $N$, but their
explicit values depend also on $v$. This means that it is important to know
which frame is chosen to be the frame of ``fiducial'' observers, in which the
observers who measure $N_{s}$ and $N_{t}$ are at rest$.$ As seen from
(\ref{itf}), (\ref{ie}) and (\ref{lt}), (\ref{cn}) the same conclusions refer
also to the determination of $E$, $B$ and $M_{s}$, $M_{t}$ from $F$ and $M$ respectively.

First, it will be assumed that $S^{\prime}$, the rest frame of the charge $Q $
is the $\gamma_{0}$-system, i.e., the frame of ``fiducial'' observers. Hence,
in $S^{\prime}$ ($\{\gamma_{\mu}^{\prime}\}$ basis), $u_{Q}=c\gamma
_{0}^{\prime}=v$, and the velocity 1-vector of the charge $q$ at any
$t^{\prime}$ is $u=u^{\prime\mu}\gamma_{\mu}^{\prime}$, where $u^{\prime\mu
}=(u^{\prime0},u^{\prime1},u^{\prime2},0)$. The results for 1-vectors $N_{s}$
and $N_{t}$ can be simply obtained using (\ref{n1}) which yields that the
``space-space'' torque $N_{s}$ as a CBGQ in $S^{\prime}$ is
\begin{equation}
N_{s}=N_{s}^{\prime\mu}\gamma_{\mu}^{\prime}=0, \label{0s}%
\end{equation}
and the ``time-space'' torque $N_{t}$ as a CBGQ in $S^{\prime}$ is
\begin{equation}
N_{t}=N_{t}^{\prime1}\gamma_{1}^{\prime}+N_{t}^{\prime2}\gamma_{2}^{\prime
}=N^{\prime01}\gamma_{1}^{\prime}+N^{\prime02}\gamma_{2}^{\prime}, \label{0c}%
\end{equation}
where $D=kQ/r^{\prime3}$ and $N^{\prime01}$, $N^{\prime02}$ are given by
(\ref{nc}).

The same results for $N_{s}$ and $N_{t}$ in $S^{\prime}$ can be obtained using
Eqs. (\ref{n2}) and (\ref{ec}), but written in terms of CBGQs. In that case
$B$ and $E$ from (\ref{ec}) are
\begin{equation}
B=B^{^{\prime}\mu}\gamma_{\mu}^{\prime}=0,\ E=E^{\prime\mu}\gamma_{\mu
}^{\prime}=D(x^{\prime1}\gamma_{1}^{\prime}+x^{\prime2}\gamma_{2}^{\prime}).
\label{e2}%
\end{equation}
Note that the spatial components of $E$ are the same as the components of the
3D $\mathbf{E}^{\prime}$ as it must.

At $t^{\prime}=0$ and when $u^{\prime\mu}=(c,0,0,0)$ all $N^{\prime\alpha
\beta}$ are zero and consequently both $N_{s}$ and $N_{t}$ are zero not only
in $S^{\prime}$ but in the laboratory frame $S$ as well.

Let us now determine 1-vectors $N_{s}$ and $N_{t}$ as CBGQs in $S$
($\{\gamma_{\mu}\}$ basis). Relative to the $S$ frame both the charge $Q$ and
the ``fiducial'' observers are moving with velocity $u_{Q}=v=\gamma_{Q}%
c\gamma_{0}+\gamma_{Q}\beta_{Q}c\gamma_{1}$. Then $N_{s}$ and $N_{t}$ in $S$
can be obtained either directly from (\ref{n1}), or by means of the LT of the
1-vectors $N_{s}$ (\ref{0s}) and $N_{t}$ (\ref{0c}). Due to invariance of any
4D CBGQ under the passive LT the ``space-space'' torque $N_{s}$ is zero in the
laboratory frame $S$ too
\begin{equation}
N_{s}=N_{s}^{\mu}\gamma_{\mu}=0.\label{1s}%
\end{equation}
The ``time-space'' torque $N_{t}$ as a CBGQ in $S$ is
\begin{equation}
N_{t}=\gamma_{Q}\beta_{Q}N^{01}\gamma_{0}+\gamma_{Q}N^{01}\gamma_{1}%
+(1/\gamma_{Q})N^{02}\gamma_{2},\label{1c}%
\end{equation}
where $N^{01}$ and $N^{02}$ are determined by (\ref{n}).

The same $N_{s}$ and $N_{t}$ in $S$ can be found using (\ref{n2}) and
(\ref{ec}) and writing all AQs as CBGQs in the $S$ frame. $E$ and $B$ in $S$
are determined as the LT of the 1-vectors $E$ and $B$ given by (\ref{e2}) (for
the LT of $E$ and $B$ see [10]). This yields
\begin{align}
B  &  =B^{\mu}\gamma_{\mu}=0,\quad E=E^{\mu}\gamma_{\mu},\nonumber\\
E^{0}  &  =D\beta_{Q}\gamma_{Q}^{2}(x^{1}-\beta_{Q}x^{0}),\ E^{1}=E^{0}%
/\beta_{Q},\ E^{2}=Dx^{2},\ E^{3}=0. \label{h}%
\end{align}
Hence $N_{s}=N_{s}^{\mu}\gamma_{\mu}=0$ and the same $N_{t}$ in $S$ is
obtained as in (\ref{1c}).

Again, we could start with the 4D angular momentums $M_{s}$ and $M_{t}$
defined by (\ref{lt}) and (\ref{cn}) and calculate them in both frames
$S^{\prime}$ and $S$. Then using the relations $N_{s,t}=dM_{s,t}/d\tau$ one
can again find the same expressions (\ref{0s}), (\ref{0c}) and (\ref{1s}),
(\ref{1c}) for $N_{s}$, $N_{t}$ in $S^{\prime}$ and $S$ respectively.

It can be easily checked that $N_{t}$ given by (\ref{0c}) in $S^{\prime}$ is
the same 4D CBGQ as $N_{t}$ given by (\ref{1c}) in $S$, i.e., that
$N_{t}^{\prime1}\gamma_{1}^{\prime}+N_{t}^{\prime2}\gamma_{2}^{\prime}%
=N_{t}^{0}\gamma_{0}+N_{t}^{1}\gamma_{1}+N_{t}^{2}\gamma_{2}$, and it is seen
from (\ref{0s}) and (\ref{1s}) that $N_{s}$ is the same 4D CBGQ for observers
in $S^{\prime}$ and $S$. This again shows that the principle of relativity is
naturally satisfied in our ISR and that there is not any paradox.

Inserting $N_{s}$ and $N_{t}$ from (\ref{0s}), (\ref{0c}) and (\ref{1s}),
(\ref{1c}) into the relation (\ref{nls}) (written with CBGQs), we can directly
check the validity of these relations.

According to our result (\ref{1s}) the 1-vector $N_{s}$ is zero both in
$S^{\prime}$ and $S$ at any $t^{\prime}$ when the ``fiducial'' frame is the
$S^{\prime}$ frame. However $N_{t}$ is different from zero in both frames. As
already said only $N_{s}$ and $N_{t}$ taken together are equivalent to the
bivector $N$, which means that validity of the above relations can be checked
measuring \emph{all six independent components} \emph{of} $N_{s}$ \emph{and}
$N_{t}$ \emph{in both frames.} It has to be remarked that the usual 3D
$\mathbf{N}$ is connected with the three spatial components of $N_{s}$.

Note that the usual 3D rotation requires measurement of only three independent
variables. Therefore in order to test SR, e.g., by means of the Trouton-Noble
type experiments, it is not enough, as usually done, to measure three
independent parameters of the 3D rotation (i.e., three independent components
of $N_{s}$, or $M_{s}$), but also one has to measure the other three relevant
variables (i.e., three independent components of $N_{t}$, or $M_{t}$).\bigskip\medskip

\textbf{C. }$N_{s}$ \textbf{and} $N_{t}$ \textbf{as CBGQs in }$S^{\prime}$
\textbf{and }$S$ \textbf{frames. }$S$ \textbf{is }

\textbf{the frame} \textbf{of ``fiducial'' observers}\bigskip

Let us now assume that the laboratory frame $S$ is the $\gamma_{0}$-system,
i.e., the frame of ``fiducial'' observers, in which the observers who measure
$N_{s}$ and $N_{t}$ are at rest, that is, $v=v^{\mu}\gamma_{\mu}=c\gamma_{0}$,
$v^{\mu}=(c,0,0,0)$. Then from (\ref{ngs}), i.e., (\ref{l1}), it follows that
in $S$ the temporal components of the 1-vectors $N_{s}$ and $N_{t}$ are zero
and only their spatial components remain. In the laboratory frame $S$ the
charge $Q$ is moving and the components of the CBGQ $u_{Q}^{\mu}\gamma_{\mu}$
are given as $u_{Q}^{\mu}=(\gamma_{Q}c,\gamma_{Q}\beta_{Q}c,0,0)$.

The 1-vectors $N_{s}$ and $N_{t}$ will be determined either directly from
(\ref{n1}) and the above expressions for $v$ and $u_{Q}$, or by the use of the
already known expression (\ref{n}) for the bivector $N$ in $S$ and the
relation (\ref{ngs}), i.e., (\ref{l1}). This yields that
\begin{equation}
N_{s}=N_{s}^{\mu}\gamma_{\mu}=N^{12}\gamma_{3},\ N_{t}=N_{t}^{1}\gamma
_{1}+N_{t}^{2}\gamma_{2}=N^{01}\gamma_{1}+N^{02}\gamma_{2}, \label{2n}%
\end{equation}
where $N^{12}$ is from (\ref{n}) or (\ref{e}) and $N^{01}$, $N^{02}$ are from
(\ref{n}). It is visible from (\ref{2n}) that in the case when $S$ is the
frame of ``fiducial'' observers \emph{the ``space-space'' torque }$N_{s}$
\emph{is different from zero.}

In the same way as in Sec. IV B we find $N_{s}$ and $N_{t}$ using (\ref{n2})
and (\ref{ec}). Now the charge\textbf{\ }$Q$ moves in the frame of
``fiducial'' observers, the $S$ frame, which yields that both $E$ and
\emph{the magnetic field} $B$\emph{\ are different from zero.} Then
\begin{equation}
E=E^{\mu}\gamma_{\mu},\ E^{0}=E^{3}=0,\ E^{1}=D\gamma_{Q}(x^{1}-\beta_{Q}%
x^{0}),\ E^{2}=D\gamma_{Q}x^{2},\label{i}%
\end{equation}
and the magnetic field is
\begin{equation}
B=B^{\mu}\gamma_{\mu},\ B^{0}=B^{1}=B^{2}=0,\ B^{3}=(D/c)\gamma_{Q}\beta
_{Q}x^{2}=\beta_{Q}E^{2}/c.\label{mg}%
\end{equation}

The spatial components $E^{i}$ and $B^{i}$ from (\ref{i}) and (\ref{mg}) are
the same as the usual expressions for the components of the 3D vectors
$\mathbf{E}$ and $\mathbf{B}$ for an uniformly moving charge. Inserting
(\ref{i}) and (\ref{mg}) into (\ref{n2}) we again find $N_{s}$ and $N_{t}$ as
in (\ref{2n}). $N_{s}$ is different from zero since $B$ given by (\ref{mg}) is
different from zero.

Instead of expressing components of $K_{L}$ in (\ref{e}) in terms of
components of the electromagnetic field $F$ we shall now write $N_{s}$ from
(\ref{2n}) in terms of 1-vectors $E$ and $B$, which are explicitly given by
Eqs. (\ref{i}) and (\ref{mg}). Then $N_{s}$ becomes
\begin{equation}
N_{s}=N^{3}\gamma_{3}=N^{12}\gamma_{3}=(\beta_{Q}ctK_{L}^{2}+(q/c)\beta
_{Q}y(E^{\mu}u_{\mu})\gamma_{3}. \label{nw}%
\end{equation}
Remember that $E^{0}=B^{0}=0,$ Eqs. (\ref{i}) and (\ref{mg}), when $S$ is the
frame of ``fiducial'' observers. In the usual approach, e.g., [2], it is
considered that in the $S$ frame the whole physical torque is the 3D
$\mathbf{N}$, i.e., $N_{z}$, given by Eq. (7) in [2]. We see that in the 4D
spacetime the physical torque is the bivector $N$ that is given by relation
(\ref{n}) as a CBGQ in $S$ and in the $\{\gamma_{\mu}\}$ basis. When that $S$
is chosen to be the frame of ``fiducial'' observers then $N$ can be
represented by \emph{two} 1-vectors $N_{s}$ and $N_{t}$ given by (\ref{2n})
and (\ref{nw}), which are both physical and have to be determined
theoretically and \emph{experimentally}. Only when the laboratory frame $S$ is
the frame of ``fiducial'' observers the spatial components of $N_{s}$ have
some resemblance with the components of the 3D $\mathbf{N}$. However note that
in (\ref{nw}) all components are the components of the 4D quantities,
1-vectors $x,$ $u,$ $K_{L},$ $E$ and $B$, while in Eq. (7) in [2] only the
corresponding 3D vectors are involved.

Let us now determine 1-vectors $N_{s}$ and $N_{t}$ as CBGQs in $S^{\prime}$.
Relative to the $S^{\prime}$ frame the charge $Q$ is at rest $u_{Q}%
=c\gamma_{0}^{\prime}$, but the ``fiducial'' observers are moving with
velocity $v=\gamma_{Q}c\gamma_{0}^{\prime}-\gamma_{Q}\beta_{Q}c\gamma
_{1}^{\prime}$. Then $N_{s}$ and $N_{t}$ in $S^{\prime}$ can be obtained
either directly from (\ref{n1}) or by means of the LT of 1-vectors $N_{s}$ and
$N_{t}$ as CBGQs, which are given by (\ref{2n}). We find that $N_{s}$ is
different from zero not only in $S$ but in the $S^{\prime}$ frame as well
\begin{equation}
N_{s}=N_{s}^{\prime\mu}\gamma_{\mu}^{\prime}=N_{s}^{\prime3}\gamma_{3}%
^{\prime},\ N_{s}^{\prime3}=\gamma_{Q}\beta_{Q}N^{\prime02}. \label{s}%
\end{equation}
For $N_{t}$ one gets
\begin{equation}
N_{t}=N_{t}^{\prime\mu}\gamma_{\mu}^{\prime},\ N_{t}^{\prime0}=-\beta
_{Q}\gamma_{Q}N^{\prime01},\ N_{t}^{\prime1,2}=\gamma_{Q}N^{\prime
01,2},\ N_{t}^{\prime3}=0. \label{3}%
\end{equation}

The same results for $N_{s}$ and $N_{t}$ in $S^{\prime}$ can be obtained using
(\ref{n2}) and (\ref{ec}) and writing all AQs as CBGQs in the $S^{\prime}$
frame. $E$ and $B$ in $S^{\prime}$ are determined by the LT of 1-vectors $E$
and $B$ given by (\ref{i}) and (\ref{mg}) respectively. They are
\begin{align}
E  &  =E^{\prime\mu}\gamma_{\mu}^{\prime},\ E^{\prime0}=-D\beta_{Q}\gamma
_{Q}x^{\prime1},E^{\prime1}=D\gamma_{Q}x^{\prime1},E^{\prime2}=D\gamma
_{Q}x^{\prime2},E^{\prime3}=0,\nonumber\\
B  &  =B^{\prime\mu}\gamma_{\mu}^{\prime},\ B^{\prime0}=B^{\prime1}%
=B^{\prime2}=0,\ B^{\prime3}=(D/c)\gamma_{Q}\beta_{Q}x^{\prime2}=\beta
_{Q}E^{\prime2}/c. \label{4}%
\end{align}
Inserting (\ref{4}) into (\ref{n2}) the same $N_{s}$ and $N_{t}$ are found as
in (\ref{s}) and (\ref{3}). $N_{s}$ in $S^{\prime}$ is different from zero
since $B$ given by (\ref{4}) is different from zero.

Of course it can be again easily seen that $N_{s}$ ($N_{t}$) from (\ref{2n})
is equal to $N_{s}$ ($N_{t}$) from (\ref{s}) ((\ref{3})); it is the same 4D
CBGQ for observers in $S$ and $S^{\prime}$.

Inserting $N_{s}$ and $N_{t}$ from (\ref{2n}) and also from (\ref{s}) and
(\ref{3}) into the relation (\ref{nls}), which connects $N$ with $N_{s}$ and
$N_{t}$, we find the same expressions (\ref{n}) and (\ref{nc}) for $N$ in $S$
and $S^{\prime}$, and we already know that they are equal, $N$ ((\ref{nc}))
$=$ $N$ ((\ref{n})).

The above discussion shows that the explicit expressions for $N_{s}$ and
$N_{t}$ depend on the choice for the frame of ``fiducial'' observers. For
example, in the $S$ frame $N_{s}$ and $N_{t}$ are given by the relations
(\ref{1s}) and (\ref{1c}), or (\ref{2n}), when $S^{\prime}$, or $S$, are
chosen for the frame of ``fiducial'' observers. However when they are inserted
into (\ref{nls}) they will always give the same $N$.

The paradox does not appear in the considered representations for the torques
since the principle of relativity is automatically satisfied in such an
approach to SR which exclusively deals with 4D geometric quantities, i.e., AQs
or CBGQs. In the standard approach to SR [17] the principle of relativity is
postulated outside the framework of a mathematical formulation of the theory,
and, as we already discussed, it is considered that the principle of
relativity holds for the equations written with the 3D quantities. \medskip\bigskip

\textbf{V. CONCLUSIONS \bigskip}

The whole consideration exposed in previous sections strongly suggests that
the violation of the principle of relativity and the existence of the
electrodynamic paradox come from the use of the 3D quantities, as physical
quantities in the 4D spacetime, and from using their apparent transformations.
We have shown that, in the 4D spacetime, the well-defined angular momentum
with relativistically correct transformation properties is not the 3D vector
$\mathbf{L}$, with its apparent transformations (\ref{lk}), but the Lorentz
invariant 4D geometric quantities, the bivector $M$, or 1-vectors $M_{s}$ and
$M_{t}$, which are derived from $M$ and which \emph{together} contain the same
physical information as $M$. The same result refers to the 3D torque
$\mathbf{N}$ and the bivector $N$, or the torques $N_{s}$ and $N_{t}$ that
together correspond to the torque $N$. It is already proved in, e.g., [8 -
12], that the same situation exists with the 3D vectors $\mathbf{E}$ and
$\mathbf{B}$ and the 4D geometric quantities, the bivector $F$, or, derived
from it, the 1-vectors $E$ and $B$.

We hope that the results obtained in this paper will have important
consequences for all branches of physics in which the relativistic effects
have to be taken into account, particularly for classical and quantum
relativistic electrodynamics. \bigskip\medskip

\textbf{REFERENCES\bigskip}

\noindent\lbrack1] J.D. Jackson, \textit{Classical Electrodynamics} (Wiley,
New York, 1977)

2nd edn.

\noindent\lbrack2] J. D. Jackson, Am. J. Phys. \textbf{72,} 1484 (2004).

\noindent\lbrack3] O. D. Jefimenko, \textit{Retardation and Relativity} (Star
City: Electret

Scientific, 1997).

\noindent\lbrack4] O. D. Jefimenko, J. Phys. A: Math. Gen. \textbf{32,} 3755 (1999).

\noindent\lbrack5] T. Ivezi\'{c}, Found. Phys. Lett. \textbf{12}, 507 (1999).

\noindent\lbrack6] T. Ivezi\'{c}, Found. Phys. \textbf{31}, 1139 (2001).

\noindent\lbrack7] T. Ivezi\'{c}, Found. Phys. Lett. \textbf{15}, 27 (2002); physics/0103026;

physics/0101091.

\noindent\lbrack8] T. Ivezi\'{c}, Found. Phys. \textbf{33}, 1339 (2003).

\noindent\lbrack9] T. Ivezi\'{c}, hep-th/0207250; hep-ph/0205277.

\noindent\lbrack10] T. Ivezi\'{c}, Found. Phys. Lett. \textbf{18,} 301 (2005)\textbf{.}

\noindent\lbrack11] T. Ivezi\'{c}, Found. Phys. \textbf{35,} 1585
\textbf{(}2005\textbf{)}.

\noindent\lbrack12] T. Ivezi\'{c}, Found. Phys. Lett. \textbf{18,} 401 (2005)\textbf{.}

\noindent\lbrack13] T. Ivezi\'{c}, physics/0505013.

\noindent\lbrack14] T. Ivezi\'{c}, Found. Phys. Lett. \textbf{12}, 105 (1999).

\noindent\lbrack15] D. Hestenes, \textit{Space-Time Algebra }(Gordon and
Breach, New York, 1966);

\textit{Space-Time Calculus; }available at: http://modelingnts.la. asu.edu/evolution.

html; \textit{New Foundations for Classical Mechanics }(Kluwer, Dordrecht, 1999)

2nd. edn.; Am. J Phys. \textbf{71}, 691 (2003).

\noindent\lbrack16] C. Doran, and A. Lasenby, \textit{Geometric algebra for
physicists }

(Cambridge University, Cambridge, 2003).

\noindent\lbrack17] A. Einstein, Ann. Physik\textit{ }\textbf{49,} 769 (1916),
tr. by W. Perrett and G.B.

\noindent\lbrack18] F. Rohrlich, Nuovo Cimento B \textbf{45}, 76 (1966).

Jeffery, in \textit{The Principle of Relativity }(Dover, New York, 1952).

\noindent\lbrack19] A. Gamba, Am. J. Phys. \textbf{35}, 83 (1967).

\noindent\lbrack20] R.M. Wald, \textit{General Relativity} (Chicago University,

Chicago, 1984); M. Ludvigsen, \textit{General Relativity,} \textit{A Geometric
Approach }

(Cambridge University, Cambridge, 1999); S. Sonego and M.A.

Abramowicz, J. Math. Phys. \textbf{39}, 3158 (1998); D.A. T. Vanzella,

G.E.A. Matsas, H.W. Crater, Am. J. Phys. \textbf{64}, 1075 (1996).

\noindent\lbrack21] D. Hestenes and G. Sobczyk, \textit{Clifford Algebra to
Geometric Calculus}

(Reidel, Dordrecht, 1984).

\noindent\lbrack22] A. Einstein, Ann. Physik \textbf{49,} 769 (1916), tr. by
W. Perrett and G.B.

Jeffery, in \textit{The Principle of Relativity }(Dover, New York, 1952).

\noindent\lbrack23] W.G.W. Rosser, \textit{Classical Electromagnetism via
Relativity} (Plenum,

New York, 1968)
\end{document}